\documentclass[reprint,superscriptaddress]{revtex4-1}

\usepackage{graphicx}
\usepackage{dcolumn}
\usepackage{bm}
\usepackage{float}

\usepackage{xcolor}
\usepackage{soul}

\newcommand{\un}[1]{\mathrm{\:#1}}

\begin{document}

\title{Robust kHz-linewidth distributed Bragg reflector laser with optoelectronic feedback}

\author{Megan Yamoah}
\email{myamoah@mit.edu}
\affiliation{Department of Physics, MIT-Harvard Center for Ultracold Atoms and Research Laboratory of Electronics, Massachusetts Institute of Technology, Cambridge, MA 02139, USA}
\author{Boris Braverman}
\email{bbraverm@uottawa.ca}
\affiliation{Department of Physics, MIT-Harvard Center for Ultracold Atoms and Research Laboratory of Electronics, Massachusetts Institute of Technology, Cambridge, MA 02139, USA}
\affiliation{Department of Physics and Max Planck Centre for Extreme and Quantum Photonics,	University of Ottawa, 25 Templeton Street, Ottawa, ON, Canada K1N 6N5}
\author{Edwin Pedrozo-Pe\~{n}afiel}
\affiliation{Department of Physics, MIT-Harvard Center for Ultracold Atoms and Research Laboratory of Electronics, Massachusetts Institute of Technology, Cambridge, MA 02139, USA}
\author{Akio Kawasaki}
\affiliation{Department of Physics, MIT-Harvard Center for Ultracold Atoms and Research Laboratory of Electronics, Massachusetts Institute of Technology, Cambridge, MA 02139, USA}
\affiliation{W.W. Hansen Experimental Physics Laboratory and Department of Physics, Stanford University, Stanford, CA 94305, USA}
\author{Bojan Zlatkovi\'c}
\affiliation{Department of Physics, MIT-Harvard Center for Ultracold Atoms and Research Laboratory of Electronics, Massachusetts Institute of Technology, Cambridge, MA 02139, USA}
\affiliation{Institute of Physics Belgrade, University of Belgrade, Pregrevica 118, 11080 Belgrade, Serbia}
\author{Vladan Vuleti\'c}
\email{vuletic@mit.edu}
\affiliation{Department of Physics, MIT-Harvard Center for Ultracold Atoms and Research Laboratory of Electronics, Massachusetts Institute of Technology, Cambridge, MA 02139, USA}

\begin{abstract}
	We demonstrate a combination of optical and electronic feedback that significantly narrows the linewidth of distributed Bragg reflector lasers (DBRs). We use optical feedback from a long external fiber path to reduce the high-frequency noise of the laser. An electro-optic modulator placed inside the optical feedback path allows us to apply electronic feedback to the laser frequency with very large bandwidth, enabling robust and stable locking to a reference cavity that suppresses low-frequency components of laser noise. The combination of optical and electronic feedback allows us to significantly lower the frequency noise power spectral density of the laser across all frequencies and narrow its linewidth from a free-running value of $1.1 \un{MHz}$ to a stabilized value of $1.9 \un{kHz}$, limited by the detection system resolution. This approach enables the construction of robust lasers with sub-kHz linewidth based on DBRs across a broad range of wavelengths.
\end{abstract}

\date{\today}

\maketitle


\section{Introduction}

Narrow-linewidth tunable lasers are fundamental components in experiments involving spectroscopy, cooling and trapping of atoms \cite{Metcalf1999}, precision metrology \cite{Ye2008}, and quantum information science \cite{Monroe2002}, where they are used to probe transitions between atomic and molecular energy levels. Since their introduction in the 1960s, diode lasers \cite{Hall1962} have proven themselves as compact and robust sources of narrow-band laser radiation.

Linewidth is an important metric which governs lasers' applicability to precision applications such as sub-Doppler spectroscopy and laser cooling of atomic species. 
The addition of optical feedback in addition to the Fabry-P\'{e}rot laser cavity introduces narrow features to the total optical gain spectrum of the laser and results in passive narrowing of the laser spectrum.
In diode lasers, this additional optical feedback is most commonly effected using an integrated grating \cite{Wang1974}, such as in distributed feedback (DFB) lasers and distributed Bragg reflector (DBR) lasers.
A cavity entirely outside the integrated device results in an external-cavity diode laser (ECDL) design \cite{Wieman1991}.
Such an external cavity increases the optical quality factor $Q$ due to greater cavity length or finesse compared to an integrated cavity \cite{Aoyama2014}. Optical feedback from an external fiber \cite{Petermann1995} has contributed to sub-kilohertz linewidth narrowing in DBR lasers \cite{Lin2012} and traditional ECDLs \cite{Samutpraphoot2014} by adding a $\sim 1 \un{m}$ long external cavity to the laser.

Further laser frequency noise reduction can be realized through active feedback, such as servo-electronic-based stabilization to a high-finesse ultra-stable cavity \cite{Drever1983}. However, the bandwidth of the feedback needs to be significantly greater than the laser linewidth if the goal is to actually narrow the laser linewidth, rather than simply suppressing the long-term drift of the mean laser frequency. 
Current diode laser systems targeted to clock applications, where linewidth narrowing to $1 \un{Hz}$ or below is required, typically stabilize only a part of the laser output by using filtering cavities and an acousto-optical modulator as a high-bandwidth feedback element \cite{Lin2012}, due to limited feedback bandwidth and nontrivial phase relationship for electrical feedback onto the diode laser current.

\begin{figure*}[hbt]
	\centering
	\includegraphics[width=1.5\columnwidth]{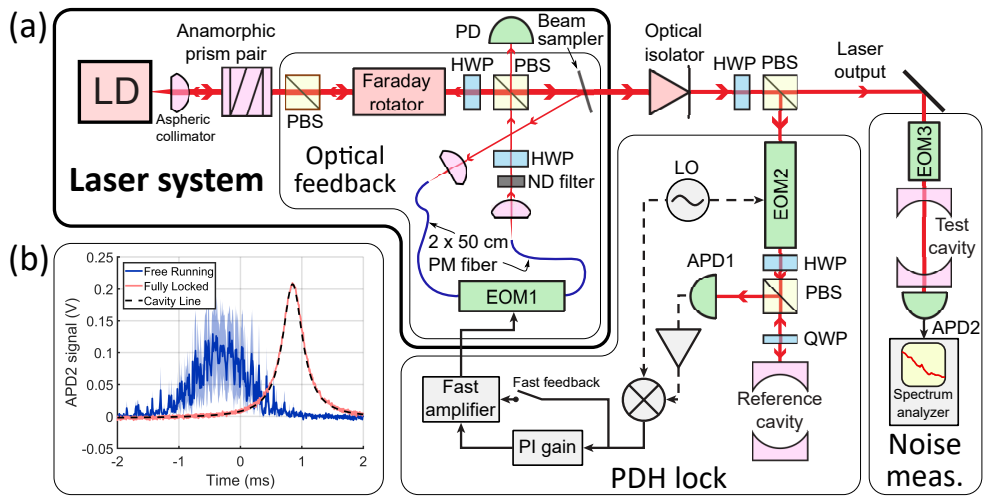}
	\caption{(a) Schematic of complete laser system with optical feedback, PDH locking, and noise characterization. The major sub-components of the setup are the laser with optical feedback through an EOM, PDH locking to the ultra-stable cavity, and the primary laser output, used to probe a test cavity and characterize the laser noise. Red solid lines represent the free-space laser beams, blue solid lines are polarization maintaining (PM) fibers, and black lines are electrical signals (dashed lines indicate RF signals). Abbreviations are: (A)PD for (avalanche) photodiode, ND for neutral density, LO for local oscillator, HWP and QWP for half- and quarter- waveplate respectively, PBS for polarizing beamsplitter.
		(b) A qualitative comparison of the laser noise between the free-running and fully locked configurations obtained by scanning the laser line across the $1.31 \un{MHz}$ linewidth test cavity at a rate of $3.3 \un{GHz/s}$.
	}
	\label{fig:ExperimentalSetup}
\end{figure*}

In the laser system presented here, we suppress the high frequency noise of a DBR laser with optical feedback from an external fiber. An electro-optic modulator (EOM) in the feedback path allows precise tuning of the laser frequency over several free-spectral ranges (FSRs) of the external cavity.
We control low frequency laser noise with a servo-electronic feedback system that locks the laser to a stable reference cavity \cite{Heinz2015}.

Our design combines passive and active linewidth narrowing schemes, resulting in a simple and robust approach to simultaneously narrow and stabilize the frequency of a DBR laser, by placing a fiber-coupled EOM within a long external feedback path. While stabilization systems for fiber lasers have included an intra-cavity EOM \cite{Hudson2005}, similar attempts for DBR or DFB lasers utilize confocal \cite{Lewoczko-Adamczyk2015}, whispering gallery \cite{Liang2015}, or fiber Fabry-Perot cavities \cite{Wei2016}.
In our system, the laser itself is completely passively running, with no manipulation of the laser diode temperature or current, allowing applying this technique to laser systems without such tunability. Moreover, since only a small fraction of the laser power (less than $-30 \un{dB}$) is used in the feedback path, the usable laser power can be kept at a large fraction of its total power, and losses within the EOM do not limit the total laser power.

We lock the laser to a cavity of $2.2 \un{MHz}$ linewidth and are able to reduce the linewidth of the DBR laser diode from a free-running value of $1.1 \un{MHz}$ to a stabilized value of $1.9 \un{kHz}$ at an averaging time of $0.1 \un{s}$. The lock we obtain remains stable for many hours and is extremely robust to mechanical vibrations. The system is currently being used in our laboratory as a master laser for an optical trap for ytterbium atoms where its exceedingly low frequency and amplitude noise levels translate to long lifetime for atoms trapped in an optical lattice \cite{Savard1997}.
Using readily available optical components, we are able to create an effective, cost-efficient, and stable system for narrowing the linewidth of and locking a DBR laser.

\section{Laser System}

The laser system is schematically outlined in Figure \ref{fig:ExperimentalSetup}(a). The laser diode (LD) is a DBR laser operating near $760 \un{nm}$ (Photodigm, P/N PH760DBR020T8-S). Its output is collimated by an aspheric lens (Thorlabs, P/N C230TME-B, $f = 4.51 \un{mm}$) and its ellipticity is corrected by an anamorphic prism pair (Thorlabs, P/N PS881-B, aspect ratio $3.5:1$). Then, the laser beam is incident on a modified $40 \un{dB}$ optical isolator (Isowave, P/N I-80T-5L) where a half-wave plate (HWP) is inserted before the output polarizing beam splitter. This allows us to keep all beam paths in the horizontal plane. The laser beam is then incident on a beam sampler (Thorlabs, P/N BSF10-B), which picks off $\sim4\%$ of the light into the feedback path.

This optical feedback light is coupled into a polarization-maintaining (PM) single-mode fiber using an aspheric lens (Thorlabs, P/N C230TMD-B, $f = 4.51 \un{mm}$), passes through a fiber-coupled EOM (EOM1, iXblue Photonics P/N NIR-MPX800-LN0.1-P-P-FA-FA), and is free-space coupled again using an aspheric lens (Thorlabs, P/N C230TMD-B, $f = 4.51 \un{mm}$) before returning to the laser through the rejection port of the optical isolator. The total length of the fiber connected to EOM1 is $1 \un{m}$, which together with the free space length of the optical feedback path corresponds to a free spectral range of $c/L \approx 100 \un{MHz}$.

A neutral density (ND) filter with optical depth of $1.3$ and a HWP control the feedback power returning to the LD, which is monitored by a photodiode (PD). By adjusting the HWP angle, we can vary the feedback power between $-30 \un{dB}$ and $-50 \un{dB}$ of the total power emitted by the LD. Feedback power ratios greater than $-35 \un{dB}$ induce coherence collapse \cite{Lenstra1985} and multi-mode behavior in the laser, while power ratios less than $-49 \un{dB}$ do not improve laser linewidth.
We characterize our system with feedback powers between $-47 \un{dB}$ and $-35 \un{dB}$ and operate the laser for long-term locking with relative feedback power near $-38 \un{dB}$.

A majority of the laser power passes through the beam sampler and into a $60 \un{dB}$ two-stage isolator (Isowave, P/N I-80U-4L). It is then split using a HWP and PBS. Most of the power ($4.2 \un{mW}$) is transmitted and constitutes the main output of the laser, while the remainder ($2.3 \un{mW}$) is sent to a Pound-Drever-Hall (PDH) \cite{Black2001} locking system. 

For the PDH lock, we modulate the light with a fiber-coupled EOM (EOM2, EOspace, P/N PM-0S5-10-PFA-PFA-770) using a local oscillator (LO) signal at $24 \un{MHz}$ and measure the reflection signal from a stable reference cavity \cite{Heinz2015} of linewidth $\kappa_r = 2.2 \un{MHz}$ using an avalanche photodiode (APD1 in Fig. \ref{fig:ExperimentalSetup}(a)). The signal is amplified and mixed with the LO to produce the PDH error signal, which passes to a proportional-integral (PI) servo-loop lockbox. The output from the PI lockbox is optionally summed with the PDH error signal (fast feedback) using a home-built amplifier with a single-pole response. The amplifier has unity DC gain and a $-3 \un{dB}$ bandwidth of $140\un{kHz}$ for the PI lockbox output, and a DC gain of $40 \un{dB}$ with $-3 \un{dB}$ bandwidth of $1.4\un{kHz}$ for the fast feedback path.

\section{Laser noise measurement}

The performance of the laser when locked to the fixed-frequency reference cavity is characterized by sending the laser output through another fiber-coupled EOM (EOM3, EOspace, P/N PM-0K5-10-PFA-760). We set the modulation frequency of EOM3 to create a laser sideband near resonance with a fixed-frequency test cavity of linewidth $\kappa_t = 1.31 \un{MHz}$. We ramp the modulation frequency to scan the laser sideband across the test cavity resonance, as shown in Figure \ref{fig:ExperimentalSetup}(b) for the free-running and fully locked configurations. The narrowing of the laser line is immediately apparent even from these qualitative data.

\begin{figure}[hbtp]
	\centering
	\includegraphics[width=0.8\columnwidth]{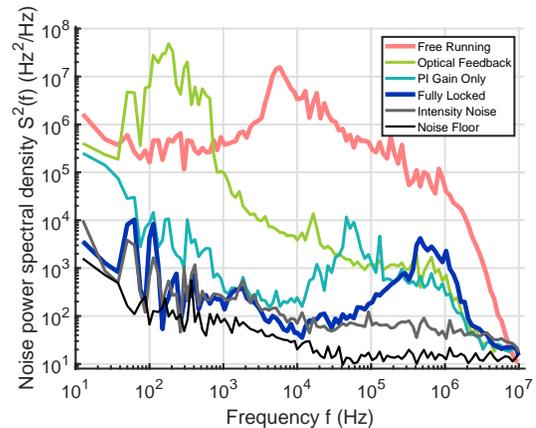}
	\caption{Laser frequency noise spectra in different operation conditions. ``Optical Feedback'' indicates configuration with optical feedback to the laser but without locking to the reference cavity. ``PI Gain Only'' refers to locking with only the servo-electronic signal while ``Fully Locked'' setup enables the ``direct feedback'' path shown in Fig. \ref{fig:ExperimentalSetup}(a).}
	\label{Fig:NoiseSpectra}
\end{figure}

To measure the laser noise power spectrum, we fix the modulation frequency of EOM3 such that the laser sideband has a detuning from cavity resonance equal to $\kappa_t/2$.
Probing the test cavity on its slope converts laser frequency noise into intensity noise, which we detect using APD2, a $20 \un{MHz}$ bandwidth APD, and a spectrum analyzer. Using the linewidth of the test cavity and the amplitude of the transmission on resonance, we can convert APD2 voltage noise into the power spectra of the laser frequency noise, as shown in Fig. \ref{Fig:NoiseSpectra}, with the conversion factor between APD2 voltage and laser frequency typically equal to $7 \un{MHz/V}$.

We characterize the laser noise in different operation regimes over the frequency range of $10 \un{Hz} < f < 10 \un{MHz}$. The ``Free Running'' configuration corresponds to the laser diode with the optical feedback path completely blocked. To keep the laser frequency from drifting away from the test cavity, the laser is loosely locked to the reference cavity using current modulation with a bandwidth $\sim 1\un{kHz}$. In all other situations, the laser current is fixed, and the voltage applied to EOM1 is the only feedback mechanism for the laser frequency.

Next, we unblock the optical feedback path and set the ratio of the feedback to total laser power equal to $-38 \un{dB}$. To observe the effect of optical feedback alone, we attenuate the PDH locking signal such that the electronic lock bandwidth is reduced to $\sim 100\un{Hz}$, yielding the curve labeled ``Optical Feedback'' in Fig. \ref{Fig:NoiseSpectra}. We see that the optical feedback reduces the high-frequency components of laser noise above $3\un{kHz}$ by approximately $25 \un{dB}$, in agreement with previous results \cite{Lin2012,Aoyama2018}. However, this comes at a price: the laser noise increases significantly below $1\un{kHz}$ because the length of the external fiber-based cavity is easily affected by low-frequency acoustical noise.

Next, we enable the servo-loop lockbox without the fast feedback, producing the ``PI Gain Only'' curve in Fig. \ref{Fig:NoiseSpectra}. PI feedback effectively suppresses low-frequency noise below the feedback bandwidth of $30 \un{kHz}$. Adding the fast feedback, we obtain the ``Fully Locked'' curve, which extends the unity-gain bandwidth out to $250\un{kHz}$. Even more notably, we see a near-complete suppression of noise across the entire frequency range between $10\un{Hz}$ and $100 \un{kHz}$ down to a level consistent with intensity noise alone.

The ``Intensity Noise'' curve is obtained under the same condition as the ``Fully Locked'' data, but by probing the test cavity on its peak rather than on its slope, which ensures that the noise observed by APD2 is entirely due to intensity fluctuations of the laser. Based on these measurements, we obtain the portion of the frequency noise spectra that can be ascribed purely to intensity noise. Finally, the ``Noise Floor'' data are taken by fully blocking the laser output going to the test cavity, and converting the data from the spectrum analyzer into equivalent frequency noise. The noise floor lies below the intensity noise of the laser across the entire frequency range and hence does not affect the measurement significantly.

\begin{figure}[hbtp]
	\centering
	\includegraphics[width=0.8\columnwidth]{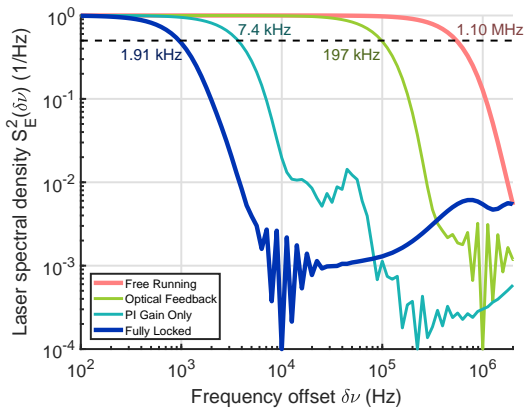}
	\caption{Laser lineshapes derived from noise power spectra shown in Fig. \ref{Fig:NoiseSpectra}. Note the servo bumps on the ``PI Gain Only'' and ``Fully Locked'' data. Numbers indicate the full width at half maximum (FWHM) of the laser line in each configuration.}
	\label{Fig:LaserLineshapes}
\end{figure}

We use the noise power spectral density measurements to calculate the laser lineshape and linewidth \cite{DiDomenico2010} at an averaging time of $0.1 \un{s}$, corresponding to the smallest measured power spectral density component. We first calculate the autocorrelation function $\Gamma_{E}(\tau)$ from the noise spectral density $S_{E}(\delta \nu)$, defined as
\begin{equation}
    \Gamma_{E}(\tau) = E^2_0e^{i2\pi\nu_0\tau}\exp\left(-2\int^{\infty}_0 S_{\delta\nu}(f)\frac{\sin^2(\pi f\tau)}{f^2}df\right).
\end{equation}
The laser power spectrum (line shape) is then given by
\begin{equation}
    S_E(\nu) = 2\int^\infty_{-\infty} e^{-i2\pi\nu_0\tau} \Gamma_{E}(\tau) d\tau.
\end{equation}
Laser line shapes calculated in this way for different feedback configurations are shown in Fig. \ref{Fig:LaserLineshapes} and show the nearly three orders of magnitude reduction of the laser linewidth under the fully locked configuration as compared to the free-running DBR laser. The extremely narrow spectrum is attained despite the $2.2 \un{MHz}$ linewidth of the reference cavity and points toward the potential of our approach for simple and robust laser systems combining the benefits of DBR/DFB and fiber lasers: small linewidths, mechanical robustness, and operation at any wavelength attainable with DBR/DFB technology. 

In keeping with tradition in laser locking experiments \cite{Corwin1998}, we tested the robustness of our laser lock by hitting the optics table with a hammer about $30 \un{cm}$ away from the laser system. When the locking is fully engaged, including the fast feedback path, the laser stayed locked despite repeated impacts of the hammer on the table, briefly jumping in frequency before returning to its locked point. This is in contrast to earlier, piezoelectric-stabilized designs \cite{Lin2012,Samutpraphoot2014}, where even loud conversation or hand-clapping can cause the laser to fall out of lock.

\section{Conclusion}

We present a method for the frequency stabilization of DBR lasers through tunable optical feedback using a fiber-coupled electro-optical modulator in the feedback path, and we anticipate this approach would work equally well for DFB lasers. Our system is simple to implement, compact, highly robust to mechanical vibration and noise, and leads to a very large reduction of laser linewidth. The laser is kept completely passively running, without modulation of the laser current or temperature, contributing to system stability. Further, our system only requires a minimal fraction of the total laser power for frequency feedback, making the majority of the laser power experimentally usable. Our approach solves the problem of producing a narrow linewidth laser for any of the large range of wavelengths available for DFB and DBR lasers, and could be broadly applied in precision spectroscopy experiments.

\section*{Acknowledgements}

This research was funded by Grant No. N00014-17-1-2254 awarded by the Office of Naval Research. B.B. acknowledges the support of the Banting Postdoctoral Fellowship.

\section*{Competing interests}

The authors have been granted a patent based in part on the work described in this paper, titled ``Semiconductor Laser with Intra-Cavity Electro-Optic Modulator'', US Patent No. 10,418,783.

\bibliography{Photodigm759}
\bibliographystyle{osajnl}

\end{document}